\documentclass[twocolumn,superscriptaddress,showpacs,preprintnumbers,amsmath,amssymb,aps]{revtex4}

\usepackage{graphicx}
\usepackage{dcolumn}
\usepackage{bm}
\usepackage{longtable}
\begin{document}


\title{Brownian rectifiers in the presence of temporally 
asymmetric unbiased forces}
\author{Raishma Krishnan}
\email{raishma@iopb.res.in}
\affiliation{Institute of Physics, Sachivalaya Marg, Bhubaneswar-751005, India}
\author{Mangal C. Mahato}
\email{mangal@nehu.ac.in}
\affiliation{Department of Physics, North-Eastern Hill University, 
Shillong-793022, India}
\author { A. M. Jayannavar }
\email{jayan@iopb.res.in}
\affiliation{Institute of Physics, Sachivalaya Marg, 
Bhubaneswar 751005, India }

\begin{abstract}
Abstract: The efficiency of energy transduction 
in a temporally asymmetric rocked ratchet is studied. 
Time asymmetry favours current in one direction and suppresses 
it in the opposite direction due to which large efficiency $\sim 50 \%$ 
is readily obtained. The spatial asymmetry in the potential together with 
system inhomogeneity may help in further enhancing the efficiency. 
Fine tuning of system parameters considered leads to multiple 
current reversals even in the adiabatic regime.\\
\end{abstract}


 \maketitle
 \section{Introduction}

Brownian rectifiers or ratchets are devices that convert nonequilibrium 
fluctuations into useful work in the presence of load. 
Several physical models ~\cite{julicher,reiman,1amj,special} 
have been proposed to 
understand the nature of currents and their possible reversals 
with applications in 
nanoparticle separation devices ~\cite{special}. 
The  possibility of enhancement of efficiency with 
which these Brownian rectifiers 
convert the nonequilibrium fluctuations into useful work 
has generated much interest in this field. This, in turn, has led to 
the emergence of a separate subfield - stochastic energetics - on its 
own right~\cite{sekimoto,parrondo}. 
Using this formalism one can readily establish the 
compatibility between Langevin or Fokker-Planck formalism with 
the laws of thermodymanics thereby providing a tool 
to study systems far from equilibrium. 
With this framework one can calculate various physical quantities 
such as efficiency of energy transduction \cite{kamgawa}, energy dissipation 
(hysteresis loss), entropy (entropy production) ~\cite{rkamj}, etc.  

Most of the studies yield low efficiencies, in the 
subpercentage range, in various types of ratchets. This is due to 
intrinsic irreversibility associated with ratchet operation. 
Only fine tuning of parameters can lead to a large efficiency, 
the regime of parameters, however, being very narrow ~\cite{sokolov}. 
Recently Makhnovskii et al. ~\cite{tsong} constructed a special type 
of flashing ratchet with two asymmetric double-well periodic-potential-
states displaced by half a period. Such flashing ratchet 
models were found to be 
highly efficient with efficiency an order of magnitude higher 
than in earlier models ~\cite{sekimoto,parrondo,astu05}. The basic 
idea behind this enhanced efficiency is that even for 
diffusive Brownian motion the choice of appropriate potential 
profile ensures suppression of backward motion and hence reduction in the 
accompanying dissipation.

In the present work, we study the motion of a particle in a rocking ratchet 
rocked purposefully as to favour current in one direction but 
to suppress motion in the opposite direction. This is in similar spirit 
as in case of flashing ratchets proposed by 
Makhnovskii et al.~\cite{tsong}. This is accomplished 
by applying temporally asymmetric but unbiased periodic 
forcings ~\cite{chivalro,phylett,ai}. Interestingly, such choice of 
forcings helps in obtaining rectified currents with high efficiency 
even for spatially symmetric periodic potentials. 
Still higher efficiency is obtained 
with asymmetric potentials. The range of parameters 
of operation of such ratchets is quite wide sustaining large loads. 
In addition, frictional inhomogeneity may further enhance the efficiency. 
We also see multiple current reversals 
in the full parameter space of operation even in the adiabatic regime. 
However, multiple current reversals require fine tuning 
of the parameters.

 \section{The Model}
The Brownian motion of a particle in an inhomogeneous 
medium is described by the Langevin equation \cite{pareekmcmdan}
 \begin{equation}
 {\dot{q}} = {- {\frac{V^\prime(q)-F(t)}{\gamma(q)}}} - 
{\frac{k_BT \gamma^{\prime}(q)}{2
 {[\gamma(q)]}^2}} + {\sqrt {\frac {k_BT}{\gamma(q)}}} \xi(t),
 \end{equation}
where $\xi(t)$ is a randomly fluctuating Gaussian 
thermal noise with zero mean and correlation, 
$<\xi(t)\xi(t^\prime)>\,=\,2\, \delta(t-t^\prime)$. 
The periodic potential $V(q)=-sin(q)-(\mu/4)\, sin(2q)$. 
The parameter $\mu (-1<\mu<1)$, characterises the 
degree of asymmetry in the potential. 
The friction coefficient $\gamma(q)=\gamma_0(1-\lambda sin(q+\phi))$, 
with $0\leq \lambda < 1$ where $\phi$ is the 
phase difference. $F(t)$ is the 
externally applied periodic driving force. 
The corresponding Fokker-Planck equation \cite{riskin} is given by
\begin{eqnarray}
\frac {\partial P(q,t)}{\partial t}&=& \frac {\partial}{\partial q}
\frac{1}{\gamma(q)}\Big[k_BT \frac{\partial P(q,t)}{\partial q}\\ \nonumber
&+& 
[V^\prime(q)-F(t)]P(q,t)\Big].
\end{eqnarray} 
Since we are interested in the adiabatic limit we first obtain an expression 
for the probability current density $j$ in the presence of a 
constant external force $F_0$. The expression is given by
\begin{eqnarray}
j&=&\frac{1-exp\,[\frac{-2\pi F_0 }{k_BT}]} 
 {{\int_{0}^{2 \pi}dy I_-(y)}},
 \end{eqnarray}
 where  $I_-(y)$ is given by
 \begin{eqnarray}
 I_-(y)&=&exp\,\left[\frac{-V(y)+ F_0y}{k_BT}\right] \nonumber \\
&\int_{y}^{y+2\pi}dx &\gamma(x) exp\,\left[\frac{V(x)-F_0x}{k_BT}\right].
\end{eqnarray}
It may be noted that for $\mu=0$, $j(F_0) 
\neq -j(-F_0)$ for $\phi \neq 0\,,\,\pi$. This asymmetry 
ensures rectification of current for the rocked ratchet 
even in the presence of spatially symmetric potential. 
We assume that $F(t)$ changes slow enough, 
i.e., its frequency is smaller than any other 
frequency related to the relaxation rate 
in the problem such that the system is in a steady 
state at each instant of time. 

We consider time asymmetric ratchets with a zero mean periodic driving force 
\cite{chivalro} given by
\begin{eqnarray}
F(t)&=& \frac{1+\epsilon}{1-\epsilon}\, F_0,\,\, (n\tau 
\leq t < n\tau+ \frac{1}{2} \tau (1-\epsilon)), \\ \nonumber
    &=& -F_0,\,\, (n\tau+\frac{1}{2} \tau(1-\epsilon) < t \leq (n+1)\tau).
\end{eqnarray}

Here, the parameter $\epsilon$ signifies the temporal asymmetry 
in the periodic forcing. For this forcing in the adiabatic limit the 
time averaged current is given by ~\cite{chivalro,kamgawa}
 $<j> = \frac{1}{2}[j(F(t)) 
+ j(-F(t))]$ which is equal to  
\begin{eqnarray}
<j>=\frac{1}{2}\,(j_1 + j_2)\,,
\end{eqnarray}
with 
\begin{eqnarray}
j_1 &=& (1-\epsilon)\, j(\frac{1+\epsilon}{1-\epsilon}F_0)\,,\\ \nonumber
j_2 &=& (1+\epsilon)\,j(-F_0).
\end{eqnarray}
The input energy $E_{in}$ per unit time is given by ~\cite{kamgawa} 
\begin{eqnarray}
E_{in}=\frac{1}{2} F_0 [(\frac{1+\epsilon}{1-\epsilon})j_1-j_2].
\end{eqnarray}  
To calculate efficiency a load $L$ is applied against 
the direction of current with overall potential given by 
$V(q)=-sin(q)-(\mu/4)\, sin(2q) + q L$. The current flows against 
the load as long as the load is less than the stopping force $L_s$ beyond 
which the current is in the same direction as that of the load. 
Thus in the operating range of the load $0<\,L<\,L_s$ the 
Brownian particles move in the direction opposite to the 
load thereby storing energy. 
The average work done over a period is given by 
\begin{eqnarray}
E_{out}=\frac{1}{2} L [j_1 + j_2]\,.
\end{eqnarray}
The thermodynamic efficiency of energy transduction is 
~\cite{sekimoto,parrondo} $\eta=E_{out}/E_{in}$. 
In our present discussion all the physical quantities are taken in 
dimensionless units. In the following section we discuss the 
results of our calculation. In order to evaluate currents we use 
the method of Gaussian quadrature.

 \section{Results and Discussions}
\begin{figure}[hbp!]
 \begin{center}
\input{epsf}
\includegraphics [width=2.8in,height=1.9in] {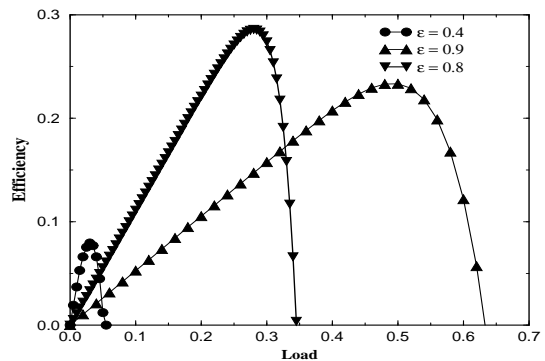}
\caption{Efficiency vs load for various values of $\epsilon$ 
with fixed $F_0 =0.1$.} \label{eff_Leps}
 \end{center}
 \end{figure}

To begin with we consider a homogeneous system 
in the presence of spatially symmetric potential. 
In Fig.~\ref{eff_Leps} 
 
\begin{figure}[htp!]
 \begin{center}
\input{epsf}
 \includegraphics[width=2.8in,height=1.9in]{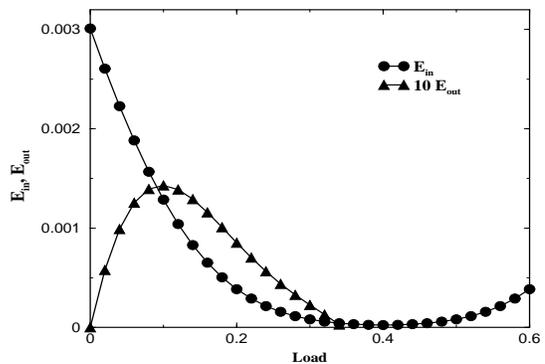}
 \caption{Input and output energy vs load for  $\epsilon=0.8$ 
with fixed $F_0=0.1$. The negative 
values of the output energy are not shown. The output curve 
is blown up ten times  to scale with the input curve values.}
\label{inoute8}
 \end{center}
 \end{figure}

\begin{figure}[htp!]
 \begin{center}
\includegraphics[width=2.8in,height=1.9in]{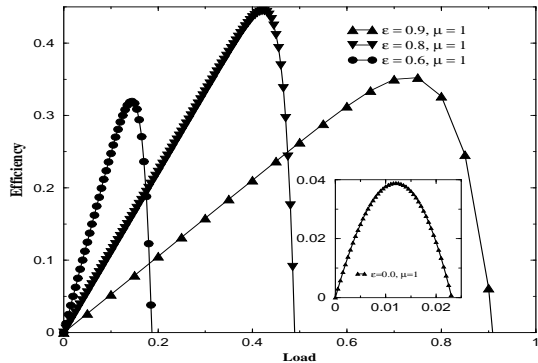}
\caption{Efficiency vs load for various 
 $\epsilon$  with fixed $F_0=0.1$. Inset shows the efficiency vs load 
 for the case $\epsilon=0$ and $\mu=1$.}\label{eff_Lem}
 \end{center}
 \end{figure}
 
\begin{figure}[hbp!]
 \begin{center}
\includegraphics[width=2.8in,height=1.9in]{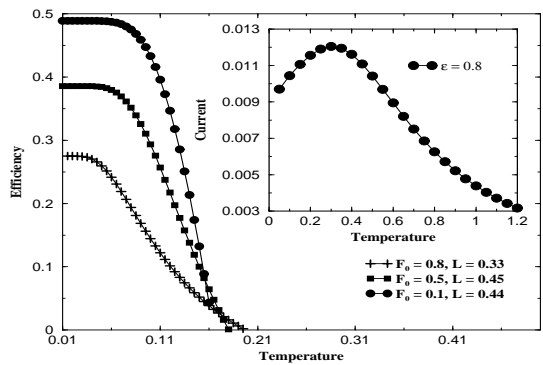}
\caption{Efficiency vs temperature with 
$\mu\,=\,1$ for (i) $\epsilon = 0.2, F_0=0.8$, (ii) $\epsilon= 0.4, F_0=0.5$ 
and (iii) $\epsilon=0.8, F_0=0.1$. Inset shows 
current as a function of temperature for 
$\epsilon\,=\,0.8, F_0=0.1$ in the absence of load.}\label{eff_T}
 \end{center}
 \end{figure}

\begin{figure}[htp!]
 \begin{center}
\input{epsf}
\includegraphics[width=2.8in,height=1.9in]{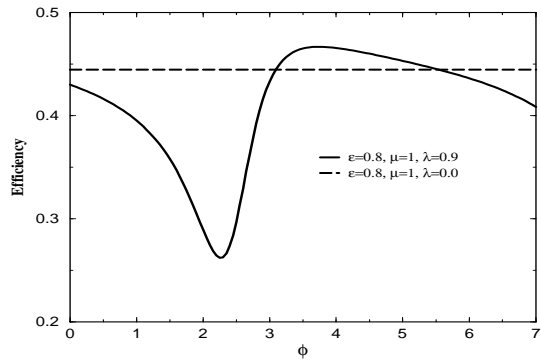} 
 \caption{Efficiency vs $\phi$ for $\epsilon\,=\,0.8, \mu\,=\,1, L=0.44$ 
 for (i) $\lambda=0.0$ and (ii) $\lambda=0.9$ 
 with fixed $F_0 =0.1$ and $T=0.1$}\label{eff_phi}
 \end{center}
 \end{figure}

\begin{figure}[htp!]
 \begin{center}
\input{epsf}
\includegraphics[width=2.8in,height=1.9in]{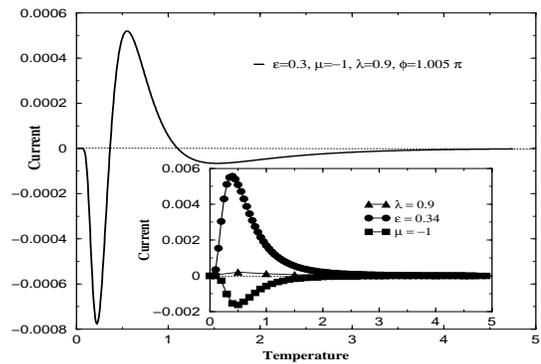}
\caption{Current vs temperature curve showing two current reversals for 
$\mu\,=\,-1, \,\lambda\,=\,0.9$ and $\epsilon\,=\,0.34$ with
$\phi\,=\,1.005 \pi$, $F_0\,=\,0.3$ and $L\,=\,0$. 
Inset shows the current in the presence of lone asymmetry parameters 
($\lambda,\epsilon,\mu$).}\label{currev2}
\end{center}
\end{figure}
\noindent we plot efficiency as a function of load 
in the presence of temporal asymmetry $\epsilon$ for 
$F_0\,=\,0.1$ and $T\,=\,0.1$. For a 
given $\epsilon$ the efficiency slowly increases with load, 
attains a maximum and then decreases rapidly. The locus of peak values 
corresponding to different $\epsilon$ values (with 
appropriate load $L$) is found to 
have a nonmonotonous behaviour with the maximum ($\sim 0.29$) being at 
around $\epsilon = 0.8$ corresponding to a load 
of $L=0.258$. 
For this value of $\epsilon$ there is a large external driving force field, 
i.e., $(\frac{1+\epsilon}{1-\epsilon}) F_0 (=9F_0=0.9)$ for a time 
duration $\tau(1-\epsilon)/2 (=0.1\tau)$, in one direction which 
causes considerable reduction in the barrier height in that direction.
This makes the particle cross the potential barrier easily 
and slide down the slope before the field changes. 
During the other part of the period of $F(t)$ $(=-F_0=-0.1)$ which 
lasts for $0.9\tau$ the potential barrier for reverse motion remains large 
thus nullifying the chances of reverse motion. There is an optimum 
value of $\epsilon$ ($\sim0.8$, for the given parameters), 
however, for which considerable current is obtained. 
Beyond this value of $\epsilon$ the duration for which the 
force remains large is so small that eventhough the barriers 
may be negligible or vanishing altogether, the particle may not have enough 
opportunity to move away before the force gets changed. This is the reason 
why the rectifier works most efficiently for an 
optimum value of $\epsilon$. For finite $\epsilon$ current 
against the load is obtained for load $L < L_s$. 
$L_s$ represents the range of load 
for which useful work is performed and is an increasing 
function of $\epsilon$.

The useful work so obtained $E_{out}$ and the input energy 
$E_{in}$ are shown in Fig.~\ref{inoute8} for a representative 
value of $\epsilon$ as a function of load. 
The input energy decreases to a minimum value 
for a load larger than $L_s$. Moreover, 
it remains positive, as expected, over the entire range. 
The output energy shows a peak with load in the region where the 
input energy is monotonously decreasing. It then becomes 
negative for $L > L_s$ as anticipated. 
The qualitative behaviours of efficiency and energies shown 
in Figs. \ref{eff_Leps} and \ref{inoute8} are similar 
to those in reference ~\cite{tsong} for the flashing ratchet. 
 
In Fig.~\ref{eff_Lem} we consider the periodic potential $V(q)$ 
to be spatially asymmetric together with a temporally asymmetric 
external driving force field. The potential 
asymmetry enhances the efficiency of energy transduction 
as well as widens the range of load. This is due to the fact that 
for $\epsilon \neq 0$ the presence of asymmetric 
parameter $\mu (>0)$ further reduces the potential 
barrier for forward motion and enhances 
the barrier for backward motion. Moreover, as can be 
seen from the inset, one can get finite current even when $\epsilon=0$ 
with finite stopping force $L_s$ in contrast to the symmetric 
potential case. From Figs. \ref{eff_Leps} and \ref{eff_Lem} it is clear 
that the temporally asymmetric forces not only enhance 
the efficiency of energy transduction but also widen the 
operation range of load against which the ratchet system works.   


In Fig.~\ref{eff_T} we plot efficiency as a 
function of $T$ for various 
$\epsilon$ values in the presence of  potential 
asymmetry ($\mu > 0$).  The efficiency decreases 
with temperature. The relevant physical 
parameters chosen for optimal efficiency are mentioned 
in the caption. From the inset it is to be 
noted that the current peaks as a function of temperature yet 
efficiency decreases monotonically. This implies that 
thermal fluctuation do not favour energy transduction 
in this case. It is worth mentioning that for 
given temperature and $\epsilon$ the efficiency shows 
peaking behaviour as a function of $F_0$; the efficiency being 
zero for $F_0=0$ as well as for large $F_0$ 
for in these limits output current vanishes in the absence of load.

Next, we present the effect of frictional 
inhomogeneity ($\gamma=\gamma(q); \lambda \neq 0$). 
In Fig.~\ref{eff_phi} we plot the efficiency as a 
function of the phase difference between the 
potential and the friction coefficient $\gamma(q)$ for a 
typical case. We observe that 
the inclusion of this parameter $\lambda$ further 
increases the efficiency in a range of $\phi$ depending on 
other parameter values. It is worth mentioning that for inhomogeneous 
systems the efficiency peaks with temperature in a limited range 
of parameters. With frictional inhomogeneity 
the range of temperature in which one can obtain output current 
with finite efficiency is extended to a large 
temperature where contribution of $\lambda$ dominates over other parameters. 
In Fig~\ref{currev2} we show that by properly choosing the parameters 
we can obtain multiple current reversals as a function of temperature. 
It should be noted that such reversals 
are not possible in the homogeneous case 
in the adiabatic regime ~\cite{danpremultiple}. 
The inset shows current as a function of individual 
parameters ($\epsilon\,,\,\mu\,,\,\lambda$).
The plots indicate that individual parameters cannot bring about 
current reversals separately. However, the possibility 
of current reversals arises 
due to the combined effect of the three asymmetry 
parameters considered. We have also observed more number of 
current reversals than shown in  Fig.~\ref{currev2} 
by fine tuning the parameters.   

\section{Conclusions}

We find large efficiency for rocking ratchets 
driven by temporally asymmetric periodic field the origin of 
which can be traced to the suppression of backward motion. 
The observed efficiency is much higher than the earlier reported 
values eventhough the ratchet operates in an intrinsically 
irreversible domain. This asymmetry factor has also helped in increasing 
the range of load of operation of the ratchet. We also observe 
multiple current reversals in the adiabatic limit by proper 
fine tuning of different parameters. These reversals are 
attributed to inherent complex dynamics of the  system.\\

\section{Acknowledgements}
AMJ thanks D.-Y. Yang for providing the reference ~\cite{tsong}
prior to publication. 
MCM thanks Institute of Physics, Bhubaneswar for hospitality where the 
present work was carried out.

 \end{document}